\catcode`\@=11
\newif\if@fewtab\@fewtabtrue
{\count255=\time\divide\count255 by 60
\xdef\hourmin{\number\count255}
\multiply\count255 by-60\advance\count255 by\time
\xdef\hourmin{\hourmin:\ifnum\count255<10 0\fi\the\count255}}
\def\ps@draft{\let\@mkboth\@gobbletwo
    \def\@oddhead{}
    \def\@oddfoot
       {\hbox to 7 cm{$\scriptstyle Draft\ version:\ \draftdate$
       \hfil}\hskip -7cm\hfil\rm\thepage \hfil}
    \def\@evenhead{}\let\@evenfoot\@oddfoot}

\def\ceqno{\global\@fewtabfalse
    \ifcase\@eqcnt \def\@tempa{& & &}\or \def\@tempa{& &}
      \or \def\@tempa{&}
      \or\def\@tempa{}\fi\@tempa
{\rm(\theequation)}}

\def\aeqno#1{\global\@fewtabfalse
    \ifcase\@eqcnt \def\@tempa{& & &}\or \def\@tempa{& &}
      \or \def\@tempa{&}
      \or\def\@tempa{}\fi\@tempa
{\rm(\theequation,#1)}}

\def\label#1{\ifnum\draftcontrol=1
 \global\def\draftnote{$\scriptstyle #1$}\fi
 \@bsphack\if@filesw {\let\thepage\relax
   \def\protect{\noexpand\noexpand\noexpand}%
\xdef\@gtempa{\write\@auxout{\string
      \newlabel{#1}{{\@currentlabel}{\thepage}}}}}\@gtempa
  \if@nobreak \ifvmode\nobreak\fi\fi\fi
  \@esphack}

\def\alabel#1#2{\label{#1}\global\@fewtabfalse
    \ifcase\@eqcnt \def\@tempa{& & &}\or \def\@tempa{& &}
      \or \def\@tempa{&}
      \or\def\@tempa{}\fi\@tempa
{\hbox to 3cm{\phantom{\rm(\theequation,#2)}
\draftnote \hfil}\hskip -3cm {\rm(\theequation,#2)}}}

\def\clabel#1{\label{#1}\global\@fewtabfalse
    \ifcase\@eqcnt \def\@tempa{& & &}\or \def\@tempa{& &}
      \or \def\@tempa{&}
      \or\def\@tempa{}\fi\@tempa
{\hbox to 3cm{\phantom{\rm(\theequation)}
\draftnote \hfil}\hskip -3cm{\rm(\theequation)}}}

\def\eqnarray{\def\draftnote{{}}\global\@fewtabtrue
\stepcounter{equation}\let\@currentlabel=\theequation
\global\@eqnswtrue
\global\@eqcnt\z@\tabskip\@centering\let\\=\@eqncr
$$\halign to \displaywidth\bgroup\@eqnsel\hskip\@centering\@eqcnt\z@
  $\displaystyle\tabskip\z@{##}$&\global\@eqcnt\@ne
  \hskip 1\arraycolsep \hfil${##}$\hfil
  &\global\@eqcnt\tw@ \hskip 1\arraycolsep
$\displaystyle\tabskip\z@{##}$
\hfil  \tabskip\@centering&\global\@eqcnt\thr@@\llap{##}\tabskip\z@
\cr}

\def\endeqnarray{\@@eqncr\egroup
      \global\advance\c@equation\m@ne$$\global\@ignoretrue}

\def\@eqnnum{\hbox to 3cm{\phantom{\rm(\theequation)} \draftnote
                         \hfil}\hskip -3cm {\rm(\theequation)}}

\def\@@eqncr{\let\@tempa\relax
    \ifcase\@eqcnt \def\@tempa{& & &}\or \def\@tempa{& &}
      \or \def\@tempa{&}
      \or\def\@tempa{}
\fi\@tempa
\if@eqnsw
\if@fewtab\@eqnnum\fi
\stepcounter{equation}\fi\global
\@eqnswtrue\global\@eqcnt\z@\global\@fewtabtrue\cr}

\def\draftcite#1{\ifnum\draftcontrol=1#1\else{}\fi}

\def\@lbibitem[#1]#2{\item{}\hskip -3cm \hbox to 2cm
{\hfil$\scriptstyle\draftcite{#2}$}\hskip
1cm[\@biblabel{#1}]\if@filesw
     {\def\protect##1{\string ##1\space}\immediate
      \write\@auxout{\string\bibcite{#2}{#1}}}\fi\ignorespaces}

\def\@bibitem#1{\item\hskip -3cm \hbox to 2cm
{\hfil $\scriptstyle\draftcite{#1}$}\hskip 1cm
\if@filesw \immediate\write\@auxout
       {\string\bibcite{#1}{\the\value{\@listctr}}}\fi\ignorespaces}

\def\draftdate{\number\month/\number\day/\number\year\ \ \ \hourmin }

\global\def\draftcontrol{0}
\catcode`\@=12

\catcode`\@=12
\def\tilde{\widetilde}
\def\hat{\widehat}
\documentstyle[12pt,epsf]{article}
\def\theequation{{\arabic{equation}}}
\newcommand{\be}{\begin{eqnarray}}
\newcommand{\en}{\end{eqnarray}\vs 0.5 cm}
\newcommand{\non}{\nonumber}
\newcommand{\no}{\noindent}

\newcommand{\Nu}{{\bf u}}

\newcommand{\NF}{{\bf F}}
\newcommand{\NU}{{\bf U}}
\newcommand{\Nx}{{\bf x}}

\newcommand{\Ny}{{\bf y}}

\newcommand{\Nk}{{\bf k}}
\newcommand{\Nl}{{\bf l}}
\newcommand{\Nna}{{\bf \nabla}}

\newcommand{\NR}{{{\bf R}}}
\newcommand{\NT}{{{\bf T}}}
\newcommand{\NZ}{{{\bf Z}}}
\newcommand{\NN}{{{\bf N}}}
\newcommand{\qq}{\begin{eqnarray}}

\newcommand{\da}{\partial}

\newcommand{\qqq}{\end{eqnarray}}

\newcommand{\tr}{\hbox{tr}}

\newcommand{\CA}{{\cal A}}

\newcommand{\hf}{{_1\over^2}}

\newcommand{\vs}{\vskip}

\begin{document}
\
\vskip 1.7cm
\begin{center}
{\large{\bf{Probabilistic estimates for the Two
Dimensional Stochastic Navier-Stokes Equations}}}

\vskip 1.4cm
J.Bricmont\\ UCL, Physique Th\'eorique,
 B-1348, Louvain-la-Neuve,
Belgium\\
\vskip 0.3cm
 A.Kupiainen\footnote{Partially supported by
EC grant FMRX-CT98-0175},
R.Lefevere \\ Helsinki University,
Department of Mathematics,\\ P.O.Box 4, Helsinki 00014,
Finland
\end{center}

\date{ }

\vskip 1.3 cm

\begin{abstract}
\vskip 0.3cm

\noindent We consider the Navier-Stokes equation on a 
two dimensional torus with
a random force, white noise in time and analytic in space, 
for arbitrary Reynolds number $R$.  We prove probabilistic 
estimates for the long time behaviour of the solutions
that imply bounds for the dissipation scale and energy
spectrum as $R\to\infty$.
\end{abstract}
\vs  1.6cm

\section{Introduction}

\vskip 0.5cm

In two dimensions global
existence and uniqueness of solutions of the
Navier-Stokes equation is known for a large class of initial
conditions and forcing, deterministic and random. In
particular, for a bounded domain the unforced system has
a finite dimensional attractor \cite{lady,cft} and this
persists for a bounded finite dimensional force.

In this paper we consider the Navier-Stokes equation with
a random force, white noise in time and large scale in
space and prove probabilistic estimates for the long time
behaviour of the solutions. Our analysis is inspired by
the recent paper by Mattingly and Sinai \cite{MS} who
gave a conceptually simple proof of analyticity of the
solutions of the 2d Navier-Stokes equation. We extend
their analysis to the random case.

We consider the stochastic Navier-Stokes equation for the
velocity field $\NU(t,\Nx)$ defined on the torus $\NT_L=
(\NR/2\pi L\NZ)^2$:
\qq
d\NU+((\NU\cdot\Nna)\NU-\nu\Nna^2\NU+\Nna p)dt=d\NF
\label{ns}
\qqq
where $\NF(t,\Nx)$ is a Wiener process with covariance
\qq
EF_\alpha(s,\Nx)F_\beta(t,\Ny)= \min\{s,t\}
C_{\alpha\beta}({_{\Nx-\Ny}\over^L})
\label{F}
\qqq
and $C_{\alpha\beta}$ is a smooth function defined on the unit
torus and satisfying $\da_\alpha C_{\alpha\beta}=0$.
(\ref{F}) represents large scale forcing, the scale
being the size of the box. (\ref{ns}) is supplemented with
the incompressibility condition $\Nna\cdot\NU=0=\Nna\cdot\NF$
and we will also assume
the vanishing averages over the torus:
$\int_{\NT_L}\NU(0,\Nx)=0=\int_{\NT_L}\NF(t,\Nx)$ which imply that $\int_{\NT_L}\NU(t,\Nx)=0$
 for all times $t$.

(\ref{ns}) implies the transport equation
for  the vorticity
$\Omega=\da_1U_2-\da_2U_1$:
\qq
d\Omega+((\NU\cdot\Nna)\Omega-\nu\Nna^2\Omega)dt=dG,
\label{ve}
\qqq
where $G=\da_1F_2-\da_2F_1$ has the covariance
\qq
EG(t,\Nx)G(s,\Ny)=L^{-2}\min\{s,t\}\Gamma({_{\Nx-\Ny}\over^L})
\non
\qqq
with $\Gamma = -\Delta\tr C$.

It is convenient to change to dimensionless variables s.t. $\nu$
and $L$ become one.  This is achieved by setting
$$
\NU(t,\Nx)={_\nu\over^{L}}\Nu({_\nu\over^{L^2}}t,{_1\over^{L}}\Nx)\, ,\,
\Omega(t,\Nx)={_\nu\over^{L^2}}\omega({_\nu\over^{L^2}}t,{_1\over^{L}}\Nx).
$$
Then $\Nu$ and $\omega$ live on the unit torus and satisfy
(\ref{ns}) and (\ref{ve}) with $\nu$ and $L$ replaced
by $1$, and $C$ and $\Gamma$ replaced by
$$
c={_{L^4}\over^{\nu^3}}C\, ,\,\gamma={_{L^2}\over^{\nu^3}}\Gamma\, .
$$
Going to the Fourier transform $\omega_\Nk(t)=(2\pi)^{-2}\int_{\NT_1}
e^{i\Nk\cdot\Nx}\omega(t,\Nx)d\Nx$ with $\Nk\in\NZ^2$ we may write
the enstrophy equation as
\qq
d\omega_\Nk=(-\Nk^2\omega_\Nk+\sum_{\Nl\in\NZ^2\backslash\{{\bf
0},\Nk\}}
(\Nk\times \Nl)|\Nl|^{-2}
\omega_{\Nk-\Nl}
\omega_\Nl )dt+df_\Nk
\label{ee}
\qqq
where ${\Nk\times \Nl}=k_1l_2-l_1k_2$ and $\{f_\Nk\}$ are Brownian
motions
 with ${\bar{f_\Nk}}=f_{-\Nk}$ and
$$
Ef_\Nk(s)f_\Nl(t)=\min\{s,t\}\delta_{\Nk ,
-\Nl}\,\gamma_\Nk
$$
and we have used the relation $\Nu_\Nk=i{_{(-k_2,k_1)}\over^{\Nk^{2}}}
\omega_\Nk$.

The dimensionless control parameter is
the $\omega$ injection rate,
$$
R=\hf\sum_{\Nk\in\NZ^2}
\gamma_\Nk=\hf\gamma(0)=\hf{_{L^2}\over^{\nu^3}}\Gamma(0),
$$
that plays the role
of Reynolds number in our model. We will be interested
in the turbulent region $R\to\infty$. We make the following assumption
on the noise covariance:
\qq
\gamma_\Nk\leq CRe^{-|\Nk|} .
\label{gamma}
\qqq
The coefficient of $|\Nk|$ is arbitrary, but we require
exponential decay. The physically relevant case is the
one with $\gamma_\Nk\neq 0$ only for a finite number of
$\Nk$ with $|\Nk|$ of the order of unity.

To state our main result, define the enstrophy
\qq
\Phi=\hf\sum_\Nk |\omega_\Nk|^2
\label{ens}
\qqq
and fix numbers $r>1$, $\alpha>1+r$. Consider, for
positive $D$, the norm
\qq
||\omega||_D=\sup_\Nk|\omega_\Nk||\Nk|^re^{D^{-\alpha}|\Nk|} .
\label{Dnorm}
\qqq
$D$ will vary below, but $r$ and $\alpha$ are fixed.
The factor $|\Nk|^r$
is useful technically (and was already used in \cite{MS}).

\vs 2mm

\no{\bf Theorem}. {\it  Let $||\omega(0)||_{D_0}
\leq D_0^\alpha <\infty$
 and $\Phi(0)=K<\infty$. Then,  there
exists a random function $D_t$ , $D_t<\infty$ for all $t$,
such that with probability $1$, $||\omega(t)||_{D_t}<
D_t^\alpha$. For any $t>C(\log D_0+ \log K)$, and for
$D^2>CR\log R$},
\qq
{\rm Prob}\{||\omega(t)||_D\leq D^\alpha \;\;\&\;\;
\Phi(t)\leq D^2\}
\geq 1-Ce^{-c{_{D^2}\over^R}}.
\label{bound}
\qqq

\vs 2mm

\no{\bf Remark.} Here and below, $C$ (and $c$) are sufficiently
large (small) constants, which may vary from place to place
but that are uniformly bounded as $R\to\infty$.
The theorem says that with probability one $\omega(t,\Nx)$ is analytic
for all times, the dissipation scale is (up to a logarithm)
$>R^{-\hf\alpha}$
and the energy spectrum
$$
e(k)\equiv k^{-1}\int_{S^1}d\hat{\Nk}E|\omega_{\hat{\Nk}k}|^2\leq
CR^{\tilde\alpha}k^{-(2r+1)}
$$
with $k=|\Nk|$,
where $r$ can be taken arbitrary close to $1$ and $\tilde\alpha$
arbitrary close to $1+r$.
These bounds hold for any fixed time  and also for the average
of these quantities over any fixed time interval. For example,
using Jensen's and Chebyshev's inequalities, one derives from (\ref{bound})
$$
{\rm Prob}\{{_1\over^T}\int_t^{t+T}|\omega_{\Nk}(s)|^2ds>
D^{2\alpha}k^{-2r}e^{-2D^{-\alpha}k}\}\leq Ce^{-c{_{D^2}\over^R}}.
$$ 
\vs 2mm

Let us close this section with two comments.
The first concerns the relationship
of our model to the standard 2d turbulence picture \cite{k,b}.
One considers (\ref{ns}) in infinite volume with
the forcing as we do at spatial scale $L$, but {\it not periodic},
rather, for instance, having a smooth Fourier transform with compact
support around $L^{-1}$. Then it is expected that a stationary
state for $\Omega$ emerges for which the energy spectrum
$e(k)=k^{-1}\int_{S^1}d\hat{\Nk}
\int d\Nx e^{ik\hat{\Nk}\cdot\Nx}E\Omega(\Nx)\Omega(0) $
has two scaling regimes
\qq
e(k)\propto \left\{\matrix{ k^{-3} &
\eta^{-1}>> k>>L^{-1} \cr
k^{-{_5\over^3}}& k <<L^{-1}} \right\}
\label{scale}
\qqq
refered to as the direct (enstrophy) cascade regime and
the inverse (energy) cascade regime respectively. The
scale $\eta$ is the ``viscous scale'' beyond which the
$e(k)$ decays more rapidly and it scales like $\nu^\hf$.
In particular, the total energy density $\int_0^\infty
e(k)dk$ is infinite in the stationary state.
 This means that starting with say
vanishing $\Nu$ at time zero, the energy density increases linearly with
time and for the ensuing stationary state only the
vorticity remains a well defined random field. One can
also work in finite volume like in this paper by forcing
the system in an intermediate scale $\eta<<\ell<<L$,
provided the energy is absorbed by friction acting on the
$|\Nk|\sim L^{-1}$ regime. This indeed is what one does
in experimental \cite{tab} and numerical \cite{massimo}
approaches.

In our case the absence of the friction forces the
energy to dissipate in the short scales too and the
spectrum should be different from (\ref{scale}).
Our bound above is certainly far from realistic,
but one would expect the $e(k)$ to diverge
as $R\to\infty$. It would be very interesting to get hold of the
direct and inverse cascade regimes, but certainly much more
sophisticated ideas are needed than what are used in
the present paper.

\vs 2mm

The second comment concerns the uniqueness of the
stationary state (the existence is standard and follows
from compactness and Lemma 1 below). In the case of Gaussian noise
like as we have there are two kinds of
results in the literature regarding uniqueness. In
\cite{flandoli} one proves uniqueness, provided the
noise is taken big enough in the ultraviolet, i.e. the
$\gamma_\Nk$ are taken to have a lower bound
$k^{-\alpha}$ for $\alpha$ sufficiently small. This
assures that ergodicity results from the action of the
noise. However, such a noise is not what one is
interested in the turbulence problem. The second result
\cite{m} is for a smooth noise but viscosity large
enough, i.e. in the nonturbulent regime. Then the
Laplacean is the dominant term in equation (\ref{ns}) and
the past is forgotten exponentially fast due to the
viscous damping.

In the turbulent regime of large R, the number $N$ of modes
$\omega_\Nk$ that are not explicitely damped by viscosity goes to
infinity as $R\to\infty$ (we get an upper bound $C R^{\alpha}$
for $N$). Nevertheless, in the absence of noise,
the enstrophy and thus $\omega(t)$ tends to
zero and this dissipativity should lead to
uniqueness of the stationary state provided
the noise is nonvanishing for these $N$ modes 
(in the case of bounded noise, kicked at discrete times,
uniqueness
has been recently proven by Kuksin and Shirikyan \cite{ks}).

From the physical point
of view, the rate of convergence to
the stationary state that could be obtained
solely due to the effect of the
noise would not be realistic. As $R\to\infty$, the relaxation time
due to this mechanism would presumably grow superexponentially
in $R$ while in actual fact
relaxation to stationarity should be due to the
nonlinearity and should be much faster.

\section{Transition probabilities}

Define the region
\qq
U_D=\{\omega\,|\,||\omega||_D\leq D^\alpha\,\,{\rm and}\,\,
 \Phi\leq D^2\}
\label{U}
\qqq
Then the basic proposition is

\vs 2mm

\no{\bf Proposition}. {\it Suppose $\omega(0)\in U_D$. Then
there are positive constants $A$ and $a$, independent of
$R$, such that}
\qq
{\rm Prob}\{\omega(t)\in U_{\sqrt{2e^{-t}}D}\, ,\, \forall
t,\,\, 0\leq t\leq 1\}\geq
1-AR^{2\alpha}e^{-a{_{D^2}\over^R}}
\label{prop}
\qqq

\vs 2mm

It has a rather immediate

\vs 2mm

\no{\bf Corollary}. {\it  Suppose $\omega(0)\in U_D$
and $D'>D$. Then}
\qq
{\rm Prob}\{\omega(t)\notin U_{\sqrt{2e^{-t}}D'}\; ,\; {\rm
for\; some}\; t\in [0,1]\}\leq
A{R}^{2\alpha}e^{-a{_{{D'}^2}\over^R}}
\label{sup}
\qqq

\no{\bf Proof.} Note that for $D< D'$, $U_D\subset U_{D'}$.
Thus $\omega(0)\in U_{D'}$. Now the Proposition implies
the claim.\hfill$\Box$

\vs 3mm

\no {\bf Proof of the Theorem.} Consider the Markov chain
with transition probabilities
\qq
p(\omega,U)={\rm Prob}\{\omega(1)\in U\;|\;\omega(0)=\omega\} .
\label{jb12}
\qqq
Let $U_n=U_{D_n}$ where $D_n^2=2a^{-1}R(\hf e)^{n}$ and define
\qq
p_{m,n}=\sup_{\omega\in U_m}p(\omega,U_n^c).
\label{jb13}
\qqq
Since, by definition, $\sqrt{2e^{-1}}D_{m} \leq D_n$, for
$m\leq n+1$,  the Corollary implies
\qq
p_{m,n}\leq A'e^{-(\hf e)^{n}}\equiv A'\pi_n
\label{pbound}
\qqq
 for $m\leq n+1$ and
$D^2_n>CR\log R$ (so that
$R^{2\alpha}e^{-a\frac{D^2_n}{2R}}\leq C'$, and we can
take $A'=C'A$).

By assumption, $\omega(0)\in U_N$ for any $N<\infty$ such
that $D_0$ in the theorem is less than $D_N$. Let
$p_n(t)\equiv{\rm Prob}\{\omega(t)\in U_n^c\}$. Then
\qq
p_{n}(t+1)\leq {\rm Prob}\{\omega(t)\in U_{n+1}\}p_{n+1,n}+p_{n+1}(t)
\leq p_{n+1,n}+p_{n+1}(t) .
\label{ite}
\qqq
Suppose, inductively in $t\in\NN$, that
\qq
p_n(t)\leq B\pi_n\;\;
\label{pb}
\qqq
for $n\geq N-t$.
 Then,
for $n\geq N-t-1$, (\ref{ite}), (\ref{pbound}) and
(\ref{pb}) yield
\qq
p_{n}(t+1)\leq A'\pi_n+B\pi_{n+1}= B\pi_n
\non
\qqq
provided we take $B= A'(1-e^{-(\hf e)})^{-1}$ (for $t=0$,
(\ref{pb}) holds for any $B \geq 0$). This completes the
induction and shows that, with probability one,
$\omega(t)\in U_n$, for some $n$, for all integer times.
Moreover, since (\ref{pb}) holds for all $n$ when $t
\geq N = C(\log D_0 + \log K)$, this finishes the proof
of the Theorem for integer times. The remaining times
follow from the Corollary.\hfill $\Box$.

\section{Enstrophy bounds}

We prove a probabilistic analogue of the enstrophy balance:

\vs 3mm

\no{\bf Lemma 1.} Given $\Phi(0)$, for any $t\in[0,1]$,
 $$
{\rm Prob} \{ \Phi(t)\geq D^2\}\leq Ce^{-{_c\over^R}
(e^{t}D^2-\Phi(0))}
$$

\vs 2mm

\no{\bf Proof.} Let $x(t)= 2 \lambda(t)\Phi(t) =
\lambda(t) \sum_\Nk
|\omega_\Nk|^2$. Then by Ito's formula (recall that $\sum_\Nk\gamma_\Nk
=2R$ and thus that
$\gamma_\Nk\leq 2R$, $\forall \Nk$):
\qq
{_d\over^{dt}}E[e^x]&=&
E[(\dot\lambda\lambda^{-1}x-2\lambda\sum_\Nk \Nk^2
|\omega_\Nk|^2+\lambda
\sum_\Nk\gamma_\Nk+2\lambda^2\sum_\Nk
\gamma_\Nk|\omega_\Nk|^2)e^x]\nonumber \\
&\leq& E[((\dot\lambda\lambda^{-1}-2+4\lambda
R)x+2\lambda R)e^x]
\non
\qqq
where  $E$ denotes the expectation taken over the
$f_{\Nk}$'s. We used the Navier-Stokes equation
(\ref{ve}), $|\Nk|\geq 1$, and the fact that that the
nonlinear term does not contribute. Take now $\lambda(t)=
{_1\over^{8R}}e^{(t-1)}$ so that $\dot\lambda\lambda^{-1}= 1$,
$\dot\lambda\lambda^{-1}-2+4\lambda
R
\leq -\frac{1}{2}$ and $2\lambda R\leq  \frac{1}{4}$. So,
\qq
{_d\over^{dt}}E[e^x]\leq E[(\frac{1}{4}-\frac{1}{2}x)e^x]\leq
\frac{1}{2}-\frac{1}{4}E[e^x]
\non
\qqq
where the last inequality follows by using $(1-2x)e^x\leq
2-e^x$. Thus,  Gronwall's
inequality implies that:
\qq
E[e^{x(t)}]\leq e^{-\frac{t}{4}}e^{x(0)} +2\leq 3e^{x(0)}
\non
\qqq
i.e.
\qq
E[\exp(\frac{c}{R}\Phi(t) e^{t})]\leq
3\exp(\frac{c}{R}\Phi(0)),
\non
\qqq
with $c=\frac{e^{-1}}{4}$
which yields the claim by Chebycheff's
inequality.\hfill$\Box$

\vs 2mm

This implies immediately the

\vs 2mm

\no{\bf Corollary}. {\it  Let $D(t)\equiv e^{-\hf t}D$
with $D^2=\Phi(0)$, and let $t_1,\dots, t_N\in [0, 1]$. Then,
\qq
{\rm Prob} \{ \Phi(t_n)\leq \frac{3}{2}D(t_n)^2,
\forall n=1, \dots, N\}\geq
1- C Ne^{-c\frac{D^2}{R}}.
\label{Cor_enstr}
\qqq
}

\section{Proof of the Proposition}

As usual, the stochastic equation (\ref{ve}) is defined
by the integral equation,
\qq
\omega_\Nk(t)=e^{-t\Nk^2}\omega_\Nk(0)+
\int_0^t ds\,e^{(s-t)\Nk^2}\sum_{\Nl\in{\bf
Z}^2\backslash\{{\bf 0},\Nk\}}(\Nk\times
\Nl)|\Nl|^{-2}
\omega_{\Nk-\Nl}(s)\omega_\Nl(s)+z_\Nk(t).
\label{inteq}
\qqq
where $z_\Nk$ is an Ornstein-Uhlenbeck process i.e.
Gaussian with mean zero and covariance
\qq
Ez_\Nk(t)z_\Nl(s)=\delta_{\Nk,-\Nl}{_1\over^{2\Nk^2}}
(e^{-(t-s)\Nk^2}-e^{-(t+s)\Nk^2})
\gamma_\Nk
\non
\qqq
Our strategy to prove the proposition is the following.
We fix a short timestep $\tau$ depending on $D$.
By the Corollary of the previous section the enstrophy
can be assumed to satisfy the required bounds at
discrete times $t_n=n\tau$. On the
interval $[0,\tau]$ we
prove an existence and uniqueness result for
(\ref{inteq}) in Lemma 3 by
imposing a suitable condition on the smallness of the
noise term $z$. At this point, the bound
for $||\omega(t)||_{D(t)}$ will not
improve as claimed in the Proposition. However the enstrophy
stays bounded and this information allows (Lemma 4)
to improve the   $||\omega(t)||_{D(t)}$-bound.
 Repeating lemmas 3 and 4
on intervals $[t_n,t_{n+1}]$ the Proposition follows.

Let
$$\tau= \delta D^{-4\alpha}$$
where $\delta$ will be chosen below (see after (\ref{ball})),
independently on $D$. We need the following standard result on
the Ornstein-Uhlenbeck process:

\vs 2mm

\no{\bf Lemma 2}. $\forall \Nk \in \NZ^2$,
$ {\rm Prob}\{\sup_{t\in[0,\tau]}|z_\Nk(t)|\geq B\tau^{\hf}\}\leq
Ce^{-{_c\over^R}e^{|\Nk|}B^2}$

\vs 2mm

This has the following simple consequence.
Let $\CA_D$ be the event
\qq
\{z\;|\;\; \forall \Nk \in \NZ^2,
\;\sup_{t\in[0,\tau]}|z_\Nk(t)|\leq
 \tau^\hf D e^{-\frac{|\Nk|}{4}}\},
\label{AD}
\qqq
 then Lemma 2 implies
\qq
{\rm Prob}\;\CA_D\geq
1-Ce^{-c{_{D^2}\over^R}}.
\label{zbound}
\qqq

We now prove  two
lemmas. The first one, as we explained above, shows
that the solution exists and that the solution satisfies the
bounds of the Proposition
over a short time interval. For this, let $Y_D$ be the
Banach space equiped with the norm $||\cdot||_D$ and
\qq
X_D=\{\omega\in C^0([0,\tau],Y_D)\;|\;||\omega||
\equiv\sup_{t\in[0,\tau]}
||\omega(t)||_{D(t)}<\infty\}
\label{X_D}
\qqq
where
$$
D(t)=e^{-\hf t}D .
$$
Then we have,

\vs 2mm

\no{\bf Lemma 3}. {\it Let $z\in \CA_D$ and
 suppose that $||\omega(0)||_D\leq D^\alpha$ and
 that $\Phi(0)\leq \frac{3}{2}D^2$. Then the solution exists in
$X_D$ and moreover,
$$||\omega(t)||_{\sqrt{2}D(t)}\leq (\sqrt{2}D(t))^\alpha\; ,\; \Phi(t)\leq 2D(t)^2
$$
 for $t\in [0,\tau]$.
}
\vs 2mm

\no  The second lemma improves on these bounds:

\vs 2mm

\no{\bf Lemma 4}. {\it Let $z\in\CA_D$ and suppose
 that $||\omega(0)||_D\leq D^\alpha$ and that
$\Phi(t)\leq 2D(t)^2$ for $t\in [0,\tau]$. Then
 $||\omega(\tau)||_{D(\tau)}\leq  D(\tau)^\alpha$.}

\vs 2mm

\no{\bf Proof of the Proposition}. Let $t_n=n\tau$.
By the Corollary in the previous section (\ref{Cor_enstr}),
we may assume that $\Phi(t_n) \leq
\frac{3}{2} D(t_n)^2$, for all $n=1, \dots, N$, where
$N=\delta^{-1}D^{4\alpha}-1$ with  probability
\qq
1-CD^{4\alpha}e^{-c{_{D^2}\over^R}}.
\label{zbound1}
\qqq
We can thus repeat Lemmas 3 and 4 on intervals
$[t_n,t_{n+1}]$, each time with probability
(\ref{zbound}). Hence, with probability bounded from
below by (\ref{zbound1})
we deduce that $||\omega(t)||_{\sqrt 2 D(t)}\leq (\sqrt{2}D(t))^\alpha$
and $\Phi(t)\leq 2D(t)^2$ for all $t\in [0,1]$,
i.e. 
we have $\omega(t)\in U_{\sqrt{2e^{-t}}D}$ as required.
By changing $c$ and $C$, we can bound  $\frac{D^{4\alpha}}{R^{2\alpha}}$
in (\ref{zbound1}) by the exponential, call $a$ and $A$ the new constants
and obtain the claim of the Proposition. \hfill$\Box$

\vs 3mm

\no{\bf Proof of Lemma 3.} Write  equation (\ref{inteq}) as
\qq
\omega=F(\omega)
\label{fp}
\qqq
where
\qq
F_\Nk(v)\equiv \omega^0_\Nk(t)+\int_0^t
ds\,e^{(s-t)\Nk^2}\sum_{\Nl\in{\bf Z}^2\backslash\{{\bf 0},\Nk\}}
(\Nk\times
\Nl)|\Nl|^{-2}
v_{\Nk-\Nl}(s)v_\Nl(s) \equiv \omega^0_\Nk(t) + N_\Nk (v)(t)
\label{map}
\qqq
and $\omega^0(t)$ equals:
\qq
\omega^0_\Nk(t)\equiv e^{-t\Nk^2}\omega_\Nk(0)+z_\Nk(t).
\label{omega0}
\qqq
Using (\ref{AD}) and $z_\Nk(0)=0$, which imply
(trivially) that $||z||
\leq D^\alpha$, and
\qq
e^{- t\Nk^2}e^{-D^{-\alpha}|\Nk|}\leq
e^{-D(t)^{-\alpha}|\Nk|},
\label{jb1}
\qqq
which holds for $t\in [0, \tau]$,
we have
\qq
||\omega^0||\leq 2D^\alpha .
\qqq
We prove now that $F$ is a contraction
in the ball
\qq
B=\{v\in X_D : ||v-\omega^0||\leq 1\},
\label{ball}
\qqq
provided the $\delta$ in $\tau=\delta D^{-4\alpha}$ is
taken small enough (independently of $D$). To show that
$F$ maps $B$ into itself, let $v\in B$. Then $||v||\leq
2D^\alpha +1$ i.e.
\qq
|v_\Nk(t)|\leq (2D^\alpha +1)e^{-D(t)^{-\alpha}|\Nk|}|\Nk|^{-r}.
\label{bound1}
\qqq
 We
must  prove that
\qq
|F_\Nk(v)- \omega^0_\Nk(t)|=|N_\Nk(v)(t)|\leq
e^{-D(t)^{-\alpha}|\Nk|}|\Nk|^{-r},
\label{bound2}
\qqq
$\forall \Nk\in {\bf Z}^2$ and
$\forall t\in[0,\tau]$.
 Inserting (\ref{bound1})
 and $|\Nk\times \Nl|\, |\Nl|^{-2}\leq |\Nk||\Nl|^{-1}$
 in the second term of (\ref{map}), we get:
\qq
|N_\Nk(v)(t)|\leq  (2D^\alpha +1)^2
\int_0^t ds\,e^{(s-t)\Nk^2}\sum_{\Nl\in{\bf
Z}^2\backslash\{{\bf 0},\Nk\}}
e^{-D(s)^{-\alpha}|\Nk-\Nl|}
e^{-D(s)^{-\alpha}|\Nl|}|\Nk-\Nl|^{-r}
|\Nl|^{-r-1}|\Nk|
\label{jb2}
\qqq
Then, using  the bound
\qq
\sum_{\Nl\in{\bf Z}^2\backslash\{{\bf 0},\Nk\}}
|\Nk-\Nl|^{-r}|\Nl|^{-r-1}\leq C|\Nk|^{-r},
\label{bound2a}
\qqq
(since $r>1$), the triangle inequality $-|\Nk-\Nl|-|\Nl|\leq
-|\Nk|$ and
\qq
\hf (s-t) \Nk^2\leq (e^{\hf \alpha s}-
e^{\hf \alpha t})|\Nk|D^{-\alpha}=
(D(s)^{-\alpha}-D(t)^{-\alpha})|\Nk|,
\label{bound3b}
\qqq
which holds for $0\leq s\leq t\leq 1$ and $D$ large
enough, one gets that
\qq
|N_\Nk(v)(t)| &\leq& (2D^\alpha
+1)^2|\Nk|C|\Nk|^{-r}
e^{-D(t)^{-\alpha}|\Nk|} \int_0^t
ds\,e^{\hf(s-t)\Nk^2} \nonumber \\
&=& (2D^\alpha
+1)^2C|\Nk|^{-r}
e^{-D(t)^{-\alpha}|\Nk|} 2 |\Nk|^{-1}(1-e^{-\hf
t\Nk^2}).
\label{bound3}
\qqq
 Since $|\Nk|^{-1}(1-e^{-\hf t\Nk^2})\leq t^{\hf}\leq
 \delta^\hf D^{-2\alpha}$
(\ref{bound2}) follows for $\delta$ small enough (but
independent of $D$). The contractive property is proven
similarily.

Combining the fact that the solution is contained in the
ball (\ref{ball})
and the inequality $2D^\alpha +1 \leq  (\sqrt 2
e^{-\frac{t}{2}}D)^\alpha
=(\sqrt 2 D(t))^\alpha$ (which holds, since $\alpha>2$,
for $t\in [0,\tau]$ and $D$ large enough) we obtain,
\qq
||\omega(t)||_{\sqrt 2 D(t)}\leq ||\omega(t)||_{D(t)}\leq
(\sqrt 2 D(t))^\alpha
\qqq
for $t\in [0,\tau]$ and $D$ large enough.

To conclude we need to
prove that $\Phi(t)
\leq 2D^2$. By (\ref{fp}, \ref{map}),
\qq
\Phi(t) = \hf \|\omega(t)\|^2_2 \leq \hf (\|\omega^0(t)\|_2 +
\|N(\omega)(t)\|_2)^2
\label{jb3}
\qqq
By (\ref{AD}) and $z_\Nk(0)=0$, the $L^2$-norm of $z(t)$
is bounded by $C\tau^\hf D=C\delta^\hf D^{1-2\alpha}$, the $L^2$-norm of
  the first term in
(\ref{omega0})
is bounded by
 $||\omega (0)|| \leq \sqrt {2\Phi(0)} \leq \sqrt 3 D$ and, using
(\ref{bound2}), the $L^2$-norm of $N(\omega)(t)$ is
bounded by $(\sum_{\Nk\in{\bf Z}^2\backslash\{{\bf 0}\}}
|\Nk|^{-2r})^\hf = C$ (since $r>1$). Thus, we obtain
the claim provided $D$ is large
enough. \hfill$\Box$

\vs 3mm

\no{\bf Proof of Lemma 4.} We note first that $\Phi(\tau)\leq
2D(\tau)^2$ implies
\qq
|\omega_\Nk(\tau)|\leq
 \sqrt 2 D(\tau)\leq D(\tau)^\alpha e^{-D(\tau)^{-\alpha}|\Nk|}|\Nk|^{-r}
\label{small_k}
\qqq
provided $|\Nk|\leq D^\beta$, $\alpha>1+r\beta$ and $D$
is large enough. Hence, we only need to consider
$|\Nk|>D^\beta$. Below, we take as $\beta$ any number
strictly larger than $1$.

\vs 3mm

We can now conclude the proof of the Lemma by using the
 following bound on the nonlinear term
of the Navier-Stokes equation,
which improves (\ref{bound3}) for $\Nk$ large enough:
\vs 3mm

\no{\bf Lemma 5}.
{\it $\forall \Nk$ such that $|\Nk|\geq D^\beta$,
and $\forall t \in [0,\tau]$,
\qq
|N_\Nk(\omega)(t)|\leq c (1- e^{-\hf
t\Nk^2}) D^\alpha
e^{-D(t)^{-\alpha}|\Nk|}|\Nk|^{-r}
\label{bnon}
\qqq
where $c$ can be taken small if $D$ is large enough.
}

\vs 3mm
Returning to the proof of Lemma 4, we have to prove
 the following bound:
\qq
|\omega_\Nk(\tau)|\leq
 e^{-\hf\alpha \tau} D^\alpha
 e^{-D(\tau)^{-\alpha}|\Nk|}|\Nk|^{-r}.
\label{jb8}
\qqq
 We shall use (\ref{map}, \ref {omega0}) and bound
each term. Using $||\omega (0)||_D\leq D^\alpha$ and
\qq
e^{-\hf \tau\Nk^2}e^{-D^{-\alpha}|\Nk|}\leq
e^{-D(\tau)^{-\alpha}|\Nk|},
\label{bnon0}
\qqq
which is similar to (\ref{jb1}),
 we get,
\qq
|e^{- \tau \Nk^2}\omega_\Nk(0)|\leq e^{-\hf \tau \Nk^2}
 D^\alpha e^{-D(\tau)^{-\alpha}|\Nk|}|\Nk|^{-r}.
\label{jb6}
\qqq
For $z_\Nk(\tau)$, use (\ref{AD}), $z_\Nk(0)=0$, and
\qq
e^{-\frac{|\Nk|}{8}}\leq
e^{-D(\tau)^{-\alpha}|\Nk|}|\Nk|^{-r}
\label{bnon1}
\qqq
for $|\Nk|$ large, to get
\qq
|z_\Nk(\tau)|
\leq \tau^\hf D e^{-\frac{|\Nk|}{8}}
e^{-D(\tau)^{-\alpha}|\Nk|}|\Nk|^{-r}.
\label{jb4}
\qqq
Finally, we use (\ref{bnon}) to bound $N_\Nk(\omega)(\tau)$.
Combining (\ref{jb6}), (\ref{jb4}), (\ref{bnon}), we obtain
(\ref{jb8}) using
\qq
e^{-\hf \tau \Nk^2}+ c (1- e^{-\hf \tau\Nk^2})+\tau^\hf
D^{1-\alpha}e^{-\frac{|\Nk|}{8}}
\leq
e^{-\hf\alpha \tau}.
\label{jb9}
\qqq
Since $\tau= \delta D^{-4\alpha}$, this last estimate
 holds for  $c$  small,
 $|\Nk|\geq D^\beta$,
and $D$  large enough.

\vs 3mm

\no{\bf Proof of Lemma 5.} Consider first the case
$D^\beta\leq |\Nk|\leq A D^\alpha$, where $A$ is a large
enough constant (chosen below). We bound
$
|\Nk\times
\Nl||\Nl|^{-2}
\leq |\Nk||\Nl|^{-1}$
and split the sum in (\ref{map}) into
\qq
(\sum_{{\bf 0}\neq|\Nl|
\leq\frac{|\Nk|}{2}}+\sum_{\Nl\neq \Nk,|\Nl|>\frac{|\Nk|}{2}})
|\omega_{\Nk-\Nl}(s)||\omega_\Nl(s)||\Nk||\Nl|^{-1}\equiv \Sigma_1+
\Sigma_2 .
\label{jb10}
\qqq
In the first sum, we bound, using Lemma 3,
$$
|\omega_{\Nk-\Nl}(s)| \leq CD^\alpha |\Nk-\Nl|^{-r}
\leq CD^\alpha |\Nk|^{-r}
$$
since $|\Nk-\Nl|\geq \hf |\Nk|$.
Also, from Lemma 3
\qq
||\omega(s)||_2 =
\Phi(s)^\hf \leq \sqrt 2 D(s)
\label{l2}
\qqq
so Schwartz' inequality
yields
\qq
\sum_{{\bf 0}\neq|\Nl|
\leq\frac{|\Nk|}{2}}
|\omega_\Nl(s)||\Nl|^{-1} \leq \sqrt 2 D(s)
\sum_{{\bf 0}\neq|\Nl|
\leq\frac{|\Nk|}{2}}
|\Nl|^{-2}\leq CD (\log|\Nk|)^\hf
\qqq
Combining these two bounds we get
\qq
\Sigma_1\leq CD|\Nk|(\log|\Nk|)^\hf D^\alpha
|\Nk|^{-r}.
\label{s1}
\qqq

For the second sum, we use
$|\omega_{\Nl}(s)| \leq CD^\alpha |\Nl|^{-r}$,
together with (\ref{l2})
and Schwartz' inequality to bound it by
\qq
\Sigma_2\leq C  D
|\Nk| D^\alpha (\sum_{\Nl\neq
\Nk,|\Nl|>\frac{|\Nk|}{2}}
|\Nl|^{-2(r+1)})^\hf \leq
CD|\Nk|D^\alpha
|\Nk|^{-r} .
\label{s2}
\qqq
Inserting (\ref{s1}) and (\ref{s2}) to
$N_\Nk(\omega)(t)$ and
performing the integral over time  we get the bound
\qq
|N_\Nk(\omega)(t)|&\leq&
CD|\Nk|^{-1}(\log|\Nk|)^\hf (1- e^{-t\Nk^2})D^\alpha
|\Nk|^{-r}
\nonumber
\\
&\leq&
C e^{CA} D^{1-\beta}(\log D)^\hf
 (1- e^{-\hf t\Nk^2}) D^\alpha
e^{-D(t)^{-\alpha}|\Nk|}|\Nk|^{-r}
\label{bnon6}
\qqq
where we used $D^\beta\leq |\Nk|\leq A D^\alpha$
and
$$
1\leq e^{-D(t)^{-\alpha}|\Nk|}e^{CA}
$$
which holds since $|\Nk|\leq A D^\alpha$.
The claim of the Lemma follows, for
$D^\beta\leq |\Nk|\leq A D^\alpha$,
 since $D$ is assumed
 to be large enough and we choose $\beta>1$.

Consider now the case $|\Nk|>AD^\alpha$.
Using the bound
 (\ref{bound3}), we get
\qq
|N_\Nk(\omega)(t)|\leq C |\Nk|^{-1}
(1- e^{-\hf
t\Nk^2}) D^{2\alpha}
e^{-D(t)^{-\alpha}|\Nk|}|\Nk|^{-r}
\leq c (1- e^{-\hf
t\Nk^2}) D^{\alpha}
e^{-D(t)^{-\alpha}|\Nk|}|\Nk|^{-r}
\label{bnon7}
\qqq
by choosing $A$ large enough (thus, we first choose $A$
large
 so that
(\ref{bnon7}) holds
with $c$ small enough for (\ref{jb9}) to be true
 and then we choose $D$ large so that the RHS of
 (\ref{bnon6})
is bounded by the RHS of (\ref{bnon}) with $c$ small
enough).\hfill $\Box$

\vs 10mm

\no{\bf Acknowledgements}. A.K. would like to thank
K.Gawedzki, S.Kuksin and Ya.Sinai for discussions.

\end{document}